\documentclass[runningheads,a4paper]{llncs}

\usepackage{amssymb}
\usepackage{graphicx}
\usepackage{multirow}
\usepackage{enumitem}
\usepackage{booktabs}
\usepackage{array}
\usepackage{mathtools}
\usepackage{longtable}
\usepackage{subcaption}
\usepackage{url}
\urldef{\mailsa}\path|{authors}@media.mit.edu|
\newcommand{\keywords}[1]{\par\addvspace\baselineskip
\noindent\keywordname\enspace\ignorespaces#1}

\begin{document}

\mainmatter  

\title{Design Strategies for Playful Technologies to Support Light-intensity Physical Activity in the Workplace}


%
%
\author{Misha Sra%
\and Chris Schmandt}
%

\institute{MIT Media Lab\\}

%
%

\toctitle{Lecture Notes in Computer Science}
\tocauthor{Authors' Instructions}
\maketitle

\begin{abstract}
Moderate to vigorous intensity physical activity has an established preventative role in obesity, cardiovascular disease, and diabetes. However recent evidence suggests that sitting time affects health negatively independent of whether adults meet prescribed physical activity guidelines. Since many of us spend long hours daily sitting in front of a host of electronic screens, this is cause for concern. In this paper, we describe a set of three prototype digital games created for encouraging light-intensity physical activity during short breaks at work. The design of these kinds of games is a complex process that must consider motivation strategies, interaction methodology, usability and ludic aspects. We present design guidelines for technologies that encourage physical activity in the workplace that we derived from a user evaluation using the prototypes. Although the design guidelines can be seen as general principles, we conclude that they have to be considered differently for different workplace cultures and workspaces. Our study was conducted with users who have some experience playing casual games on their mobile devices and were able and willing to increase their physical activity. 

\end{abstract}

\section{Introduction}

Recent evidence suggests short bouts of physical activity (i.e., standing up, stretching) are positively associated with cardiorespiratory fitness \cite{Ross}. A basic premise of ``inactive physiology" is that sitting too much is not the same as lack of exercise and has its own metabolic consequences \cite{Hamilton}. With that in mind, we try to answer the question of what an individual who sleeps eight hours a day and exercises for 30 minutes, can do the rest of the 15.5 hours of their day they are not exercising in order to improve their health and wellbeing. Studies show what people do in their non exercise time may be equally beneficial for health as 30 minutes of recommended focused exercise activity \cite{Hamilton}. Substantial research has been devoted to promoting physical activity using mobile devices through activity trackers and sensors that calculate the number of steps walked, distance run, calories expended, or heart rate achieved. These systems employ additional approaches like virtual rewards, game elements, social incentives, and goal setting to motivate users to be more active with varying degrees of success. However, there is limited prior research on promoting physical activity \cite{Morris:2008:SUI:1357054.1357337} during work breaks and we believe this is a missed opportunity for beneficially using time to ameliorate the adverse health effects of sedentary behavior. The growing prevalence of stand-up and treadmill desks shows a desire for creatively addressing the perils of sitting. Our target audience is graduate students and staff in a research university environment. It is generally believed that good habits learned during childhood stay with us for life. Following from that belief, we hope to create new exercise habits in our peer group population so that when we graduate and juggle work and family responsibilities, we don't neglect our wellbeing  and good habits learned during graduate school stay with us. 

We believe providing a digital interface to familiar physical actions with game elements like goals, challenges, levels, rewards, and high scores could add motivation for frequent use necessary for impacting sedentary behavior. Thus, we designed three physically interactive digital games: See-Saw, a balancing game for Google Glass; Jump Beat, a music beat matching game for Google Glass; and Learning To Fly, a Microsoft Kinect game where two users fly a virtual bird through obstacles by flapping their arms in sync. In addition to whole-body movement, we capture heart rate patterns by attaching a pulse sensor to Glass to make a connection between the user's physiological state and their context of interaction. Each game is designed to explore different user motivations from free-form self directed play in See-Saw to guided play in Jump Beat to cooperative play in Learning to Fly to help create a comprehensive set of design guidelines. We evaluated prototypes of the three games with a group of users who wanted to increase their levels of physical activity. The results of the user study are being used to inform the design of new applications to enable a longitudinal study for examining impact on sedentary behavior.

In this paper, we focus our discussion on the four key design elements for technologies that encourage
physical activity which we derived from our analysis of the qualitative data collected from the user study: 1) Provide personal awareness of current activity state, 2) Use simple game mechanics for mapping physical actions to virtual elements, 3) Support social interactions, and 4) Consider the practical constraints of users' workplace culture, space, interests, and ability. The remainder of this paper is organized as follows: First, we present related work. Next, we describe the design of the three games. We follow the design with the user study methodology and discuss the results. Finally, we present the design guidelines and conclude with a summary of our findings.

\section{Related Work}

Our work is informed by prior work in three distinct areas: depth and sensor based approaches for physical interaction; movement-based play; and short duration breaks at work. Research has shown exergames, which combine exercise and digital games motivate people to be more physically active \cite{Mueller}. Our goal is to ameliorate the adverse effects of extended periods of sitting, an unavoidable consequence of modern work and lifestyles, using playful mechanisms. Increasing bodily movement by taking breaks from sitting is advocated as a means for increasing metabolism \cite{Healy} and reducing the risk of cardiovascular disease \cite{Lee}. Recent evidence suggests health consequences of ``too much sitting" are distinct from those of ``too little exercise" \cite{Hamilton}. Substantial amount of research has been devoted to promoting physical activity in general through mobile devices with embedded sensors \cite{Consolvo,Fujiki:2010:IPA:1753846.1754146} or through dedicated devices like the Nike Fuel Band\footnote{Nike. http://www.nike.com/us/en\_us/c/nikeplus-fuelband} or the FitBit\footnote{Fitbit. http://www.fitbit.com} to calculate number of steps walked, distance run, or calories expended. However, there is limited work that promotes physical activity during prolonged sitting at work and we hope to fill this gap.   

\subsubsection{Physical Interaction}

In \emph{Brave NUI World}, Wigdor and Wixon describe a natural user interface as a combination of UI and experiences created by leveraging the potential of new technologies to better mirror human capabilities \cite{mit.00210851220110101}. Physical interactions for manipulating digital information using motion-based controllers are an example of \emph{natural} interaction with technology. The recent Kinect Sports Rivals\footnote{Kinect. http://www.xbox.com/en-US/xbox-one/games/kinect-sports-rivals} is an example game that scans and creates a stylized digital version of the user. It uses body movements for interaction and reflects the player's expressions on their avatar's face. More recently, free-form hand gestural interaction with head-mounted displays (HMDs) is made possible with the Mime sensor \cite{Colaco:2013:MCL:2501988.2502042} while the Leap Motion Controller\footnote{Leap Motion. https://www.leapmotion.com/} allows hand and finger gestures as input for laptop devices. A variety of interactive systems are feasible with gestural interaction like creating animations with KinEtre \cite{Chen:2012:KAW:2380116.2380171} recording and learning physical movement sequences with YouMove \cite{Anderson:2013:YEM:2501988.2502045}, and supporting immersive video see-through augmented reality (AR) with the Oculus Rift\footnote{Oculus VR. http://www.oculusvr.com} in AR-Rift\footnote{AR-Rift. http://willsteptoe.com/post/66968953089/ar-rift-part-1}. We take inspiration from natural user interfaces in the design of gestural and whole body movement-based interaction for our games. 

\subsubsection{Movement-based Play}

Beyond encouraging social connection, play constitutes an emotionally significant context that can transform tedious physical activity into an enjoyable experience. Movement-based play puts the body in the center of the interactive experience where gross-motor bodily input influences outcomes in the digital world \cite{Edge}. Through play, exploration or other similar activities people experience positive emotions with benefits to physical, intellectual, and social well-being. \cite{aspinall2013urban,Isbister}. Though a relatively recent phenomenon, there exists evidence to suggest the potential of exergames to encourage physical activity \cite{Bianchi,Yim:2007:UGI:1328202.1328232}. Commercially available games like Dance Dance Revolution have been found to increase aerobic fitness \cite{Gao} and similarly WiiFit has been found to be more enjoyable than traditional exercise \cite{Lyons}.  Exergames in research have integrated standard exercise equipment like bikes with play \cite{Yim:2007:UGI:1328202.1328232} and used physiological sensing like player's heart-rate for controlling the game \cite{deOliveira:2008:TEE:1409240.1409268}. The design of our prototype games uses a combination whole-body actions and physiological signals as input for the digital systems. 

\subsubsection{Short Duration Breaks}

Research studies have looked at break-reminder software as a means of improving voluntary break-taking with positive effects on productivity and well-being \cite{Hennings}. Lin et al. \cite{Lin:2006:FEP:2166283.2166299} have used personal awareness and social pressure to increase daily walking while Morris et al. have used physical interactions to motivate good ergonomic habits \cite{Morris:2008:SUI:1357054.1357337} in SuperBreak. In a study of SitCoach \cite{Dantzig:2013:TPM:2559051.2559079} it was shown that persuasive messages significantly reduced computer activity, however, no significant changes in physical activity were observed. Our work focuses on the design of short duration digital games to encourage physical activity during work breaks to reduce overall sitting time. The length of short breaks mostly depends on the individual and the workplace culture though some recommendations suggest taking a three - five minutes break 25 minutes e.g. the Pomodoro technique \footnote{\url{http://pomodorotechnique.com}}.

\section{Implementation}

Our prototypes are composed of a piece of commercial hardware--Google Glass or Microsoft Kinect--and our custom software that runs on each. In this section, we describe the three applications that were used in the user study. 

\begin{figure}[t]
\centering
   \begin{subfigure}{0.49\linewidth} \centering
     \includegraphics[scale=0.2]{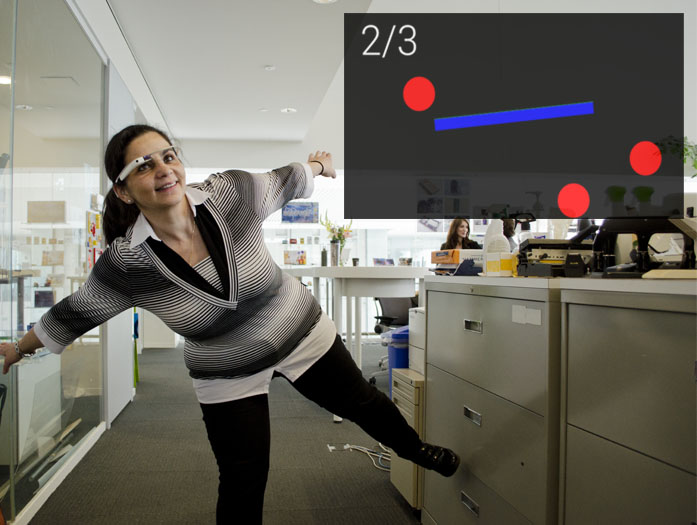}
     \caption{See-Saw on Glass.}\label{fig:seesaw}
   \end{subfigure}
   \begin{subfigure}{0.49\linewidth} \centering
     \includegraphics[scale=0.2]{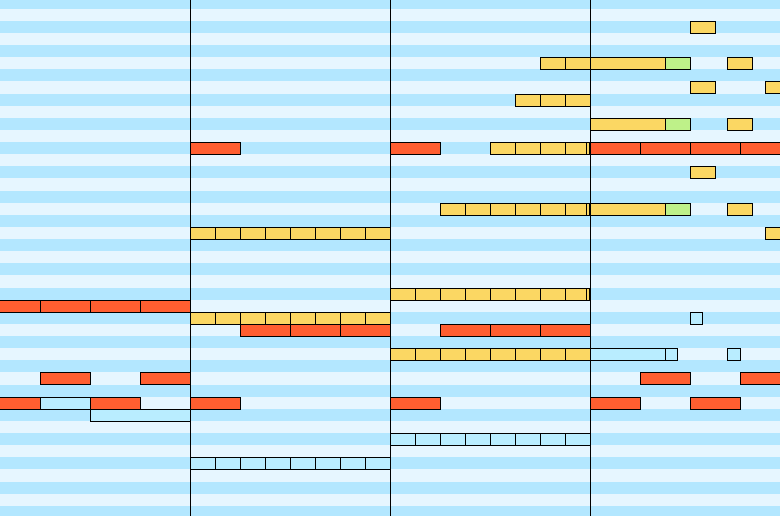}
     \caption{Jump Beat midi sequence.}\label{fig:midi}
   \end{subfigure}
\caption{\textbf{(a)}: Glass screen (inset) showing ball 3 of 3 about to fall off the beam as the user tries to balance themselves playing See-Saw. \textbf{(b)}: Partial midi sequence of a song used in Jump Beat showing drums in orange and electric bass in yellow in MidiEditor on Windows 7. } \label{fig:twofigs}
\end{figure}

\subsubsection{See-Saw on Google Glass}

In See-Saw, the user is tasked with balancing a virtual ball on a beam (see Figure~\ref{fig:seesaw}). There is no time, equipment, and overhead required to setup the system which makes it convenient for people to interrupt their day with multiple short bursts of activity by simply putting on Glass and using voice commands to start the See-Saw application. See-Saw has 2D (two dimensional) visuals and is written in Java using the Glass Development Kit (GDK) and the Android SDK. The simple design and free-form interaction allows users to challenge themselves by incorporating rules on the fly like standing on one foot, walking, or jumping while trying to balance the ball. The embedded accelerometer and gyroscope in Glass provide data for calculating the orientation of the user's head. The returned angle value is used for animating the blue beam on screen. We used the Jbox2D\footnote{A Java port of the Box2D physics engine. http://www.jbox2d.org/} physics engine for bringing the experience closer to that of balancing a real ball. The physical input from the user tilting their head left or right keeps the beam level and prevents the ball from falling off the beam. The sensitivity of the simulated physics can be adjusted to provide an easier or more challenging experience. 

\subsubsection{Jump Beat on Google Glass}

\begin{figure*}[t!]

\centering
\includegraphics[width=0.9\columnwidth]{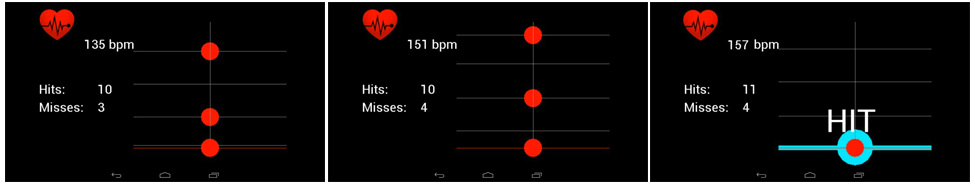}
\caption{Three Glass screenshots showing sequence of progression as user misses one scrolling note (middle screenshot) and matches the next one (right screenshot) in Jump Beat. User score and BPM visible in each screenshot.}
\label{fig:jumpbeatseq}
\end{figure*}

The game requires users to jump in time to match music notes that scroll on-screen in order to score points (see Figure~\ref{fig:jumpbeatseq}). The visuals are 2D and the game takes inspiration from an older console game Guitar Hero\footnote{Guitar Hero. http://www.guitarhero.com/}. The frequency of notes is affected by the user's heart rate which is measured by a pulse sensor. When the heart rate goes over a pre-set threshold, the frequency of notes slows down allowing the user to catch their breath and if it goes below the threshold, the frequency of notes goes up requiring the user to jump often to keep up with the game. The notes are matched to drum beats in the song and for our prototype we created one playable song. In order to match the notes on screen to the beats in the song we converted the song MP3 to midi (see Figure~\ref{fig:midi}) and manually analyzed the result for beat frequency. Based on resulting frequency we generated a map to use in our software for synchronizing the animation to the beats. 

\begin{figure}[t]
\centering
   \begin{subfigure}{0.49\linewidth} \centering
     \includegraphics[scale=0.18]{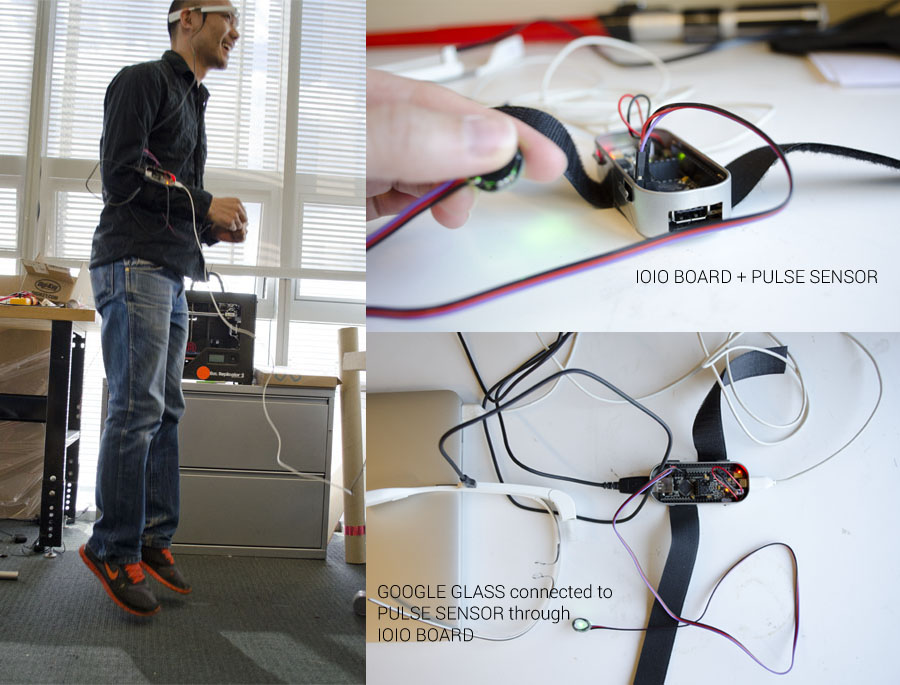}
     \caption{Jump Beat on Glass.}\label{fig:jumping}
   \end{subfigure}
   \begin{subfigure}{0.49\linewidth} \centering
     \includegraphics[scale=0.18]{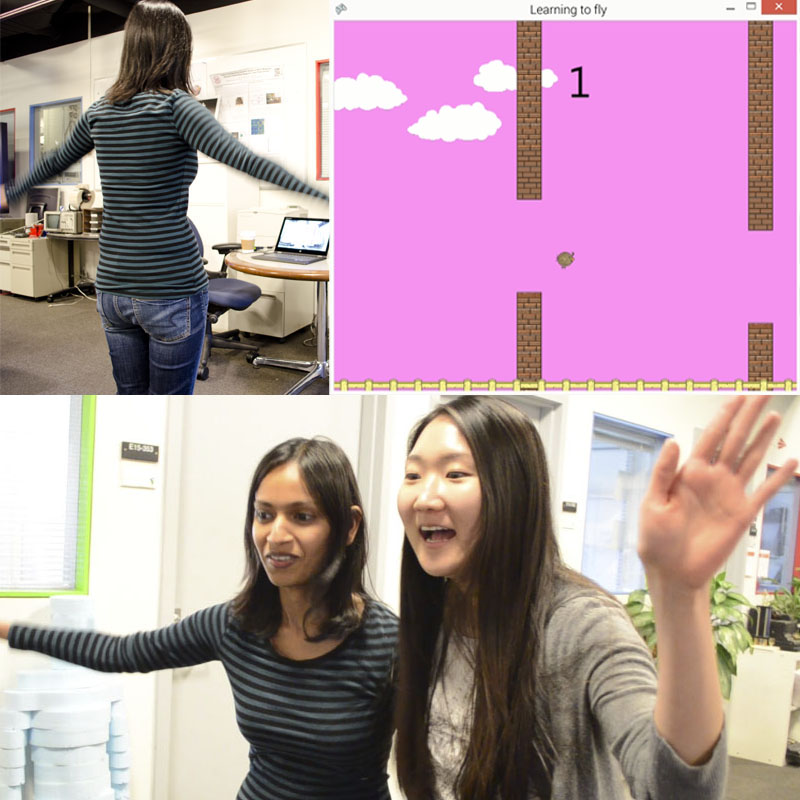}
     \caption{Learning to Fly on Kinect.}\label{fig:fly}
   \end{subfigure}
\caption{\textbf{(a)}: user jumping to match scrolling notes visible on Glass screen (see Figure ~\ref{fig:jumpbeatseq}). The pulse sensor is held between the right index finger and thumb and connect to Glass through the IOIO board. Figure (a) top right: pulse sensor connected to IOIO board. Figure (a) bottom right: Glass connected to pulse sensor through the IOIO board. \textbf{(b)}: one person playing Learning To Fly by flapping both arms. Figure (b) top right: screenshot of Learning to Fly showing player high score and bird flying through obstacles. Figure (b) bottom right: two users interacting with Learning To Fly by flapping their arms together.} \label{fig:twofigs}
\end{figure}

Data from a pulse sensor is employed for providing input into the system in addition to the ``jumping'' action by the user (see Figure~\ref{fig:jumping}). The sensor input is dependent on the ``jumping'' action which is used in two ways for affecting the system: directly as a physical interaction mechanic such that jumping is equivalent to a button press of a game controller or a key press on the keyboard; indirectly as heart rate data that affects the frequency of music notes scrolling on the screen. The heart rate calculated from the pulse sensor depends on how much the user is jumping and how the jumping is affecting them physiologically. The pulse sensor can be held between the thumb and index finger or clipped to an ear lobe to provide hands-free sensor data. We found the sensor to be more sensitive when clipped to the ear than when held in the hand for some users. Holding the pulse sensor too hard can squeeze all the blood out of the fingertip leading to poor or no signal while holding it too lightly can lead to noise from movement and ambient light. The pulse sensor\footnote{We used a commercially available pulse sensor from https://www.sparkfun.com/products/11574} is connected to Glass through the IOIO board\footnote{IOIO. https://github.com/ytai/ioio/wiki} over USB (see Figure~\ref{fig:jumping}).

\subsubsection{Learning To Fly on Microsoft Kinect}

The game was designed to invite participation from officemates or friends for a collaborative and social work break but it can also be played alone (see Figure~\ref{fig:fly}). The game is a 2D clone of a currently popular mobile game ``Flappy Bird'' but instead of touch based interaction it uses movement-based input to keep a virtual bird in flight. The onscreen bird moves from the left side of the screen to the right and encounters series of walls with openings at varying heights that it needs to go through. If it flies into a wall the game starts over. In the two player mode, both people need to flap one arm each and stay synchronized for keeping the bird in the air. The game is written in C\# using the Microsoft Kinect Toolkit and the Microsoft XNA Framework with the Farseer Physics Engine. The Kinect sensor is used to get skeleton and joint data for users standing in front of the sensor. A \emph{flap} is defined as a cycle of movement where both arms go up, down, and back up together for one user. Two players use one opposite arm each to \emph{flap} synchronously (see Figure ~\ref{fig:fly}). The position and movement over time of the hand joint in each arm, relative to the respective elbow and shoulder joints is used to calculate the \emph{flap}. 

Both hands for each user (Kinect provides tracking for a maximum of two skeletons while detecting up to 6) were independently tracked to allow the system to be flexible and usable by one person or two people. The design allows for a second player to fluidly enter or exit an ongoing one player game without requiring a restart of the game. The XNA platform provided the physics engine used to simulate gravity in the game as well as create the canvas for displaying animations. The interaction involved alternately flapping and letting go in order to guide the bird through gaps in walls that occur at varying heights.

\section{Evaluation}

We conducted an experimental user study with eleven volunteer participants (ages 25--44, 5F, mean age: 31.09, standard deviation: 6.52) to gain useful feedback from end users. Three users played only See-Saw while eight played Jump Beat and Learning to Fly. To learn about social aspects and acceptance of physical activity in open or shared workspaces we brought the devices to the users' offices instead of conducting a lab study. Participants were asked beforehand if they would comfortably be able to do the physical actions required in each game since our current system is targeted at users who are willing and able to do the physical actions. Before each session, users were introduced to the goal and the physical action required to interact with each game. Participants filled out a post-study questionnaire, with ten of the fourteen items consisting of a five-point Likert scale with two bipolar anchors to mark the opposing ends of the scale (\emph{never - frequently, frustrating - fun}). The rest of the questions asked whether they enjoyed the game, how it compared to their existing break activities, whether the physical interaction was engaging, and how it felt playing in a shared workspace. In addition to estimating the overall value of physically interactive breaks, we assessed the subjective perception of each of the three games, to inform the design of future physical interaction based systems.
Each test session comprised of participants playing the game for two to five minutes followed by 25 minutes\footnote{Based on the Pomodoro technique http://pomodorotechnique.com/} of work, during which they were encouraged to continue as they would on a normal work day. At the 25 minute mark they were asked to play again. We repeated this over time to include three breaks and four play sessions. Since for several participants, this was the first time they had used Glass, we tried to counter the novelty effect by asking participants to play multiple times. Still some of the success of See-Saw and Jump Beat may partially be attributed to the novelty of using new technology.

\subsection{Results}

Regarding their experience, 100\% found playing a game fun and engaging even though only two out of eleven admitted to playing casual games on their mobile devices frequently.  Ten participants found the amount of physical activity per session ``just right'' while one person did not think it was even noticeable. The activities were rated enjoyable with a median response of 4 and variance 0.73 on a scale of 1 to 5 where 5 meant extremely enjoyable. The physical interaction was more enjoyable than keyboard based interaction as confirmed by the Wilcoxon test\footnote{Wilcoxon Test. \url{http://en.wikipedia.org/wiki/Wilcoxon_signed-rank_test}} with p < 0.05, indicating a trend and not mere coincidence in the data. 87.5\% of the users enjoyed jumping to music, 100\% enjoyed the social interaction, and 66.66\% enjoyed improvising in See-Saw. The effect of the heart rate sensor on the game was not noticeable to most users but people enjoyed seeing their BPM (beats per minute) values. Two participants did not notice anything on the Glass display other than the scrolling notes and that may be due to the small size of the screen and the focus required to coordinate jumping with the scrolling notes. 87.5\% participants specifically mentioned enjoying the activity on Glass because of the hands-free experience. Almost all participants said they would like to integrate physical games and other activity into their break time. Six participants shared offices with one other person while one participant had two officemates. Others worked in an open space. 

\subsection{Key Design Guidelines}

Several themes emerged from our analysis of the qualitative data, which we present as four key design guidelines of technologies that encourage physical activity in the workplace:
\begin{enumerate}
\item Provide personal awareness of current activity state,
\item Use simple game mechanics for mapping physical actions to virtual elements,
\item Support social interactions, 
\item Consider the practical constraints of users' workplace culture, space, interests, and ability.
lifestyles.
\end{enumerate}

Based on user evaluation we formulated a set of design guidelines for the development of systems for introducing light-intensity physical activity into our sedentary work lives. We believe these guidelines are applicable to the design of movement-based casual game systems in general and expand upon existing casual game design guidelines \cite{deterding2011game}. Understanding game play, social interaction, and spectatorship also provide valuable input to the design of our physically interactive systems for semi-private workspaces. 

\subsubsection{Use simple game mechanics for mapping physical actions to virtual elements}

Most video games are designed for dedicated play. Games that can be played during short breaks at work need to be quick and easy to learn with fast reachable goals similar to casual games\footnote{\url{http://en.wikipedia.org/wiki/Casual_game}} on mobile devices. This is also important for multiplayer games where the game may be concurrent with other activities like social interaction. In addition, identifying the right kind of challenges for addressing different skill levels and providing an evolving experience to keep the game interesting are valuable design considerations. All participants said the games were ``quick fun" and a few remarked ``jumping and flapping felt natural." One user found the lack of explicit game mechanics in See-Saw confusing and wanted ``instructions to show on screen." Most of the users said they liked playing a physical version of a digital game they were familiar with saying they ``already know how to play" and ``I can get good at it fast." Mapping physical input with corresponding digital functions correctly is important for creating intuitive user experiences. For example, asking the user to move their head side to side in order to make a virtual bird fly would not be as natural as asking them to flap their arms. Another valid strategy would be to select familiar physical movements and connecting the action with things that are universally compelling like music in Jump Beat. Additionally, similar to traditional games, if designing a multiplayer game, it should allow people to join or leave the game at any point without adversely affecting the game experience. 

\subsubsection{Provide personal awareness of current activity state}

In movement-based interaction, the body is the major focus of attention \cite{Mueller}. The design of the visual interface should be simple and not distract from this focus on the body and provide enough information such that the user is able to make a connection between their physical action and it's virtual counterpart. The user should be able to view the visual representation on screen, consider how to affect the elements in the represented system and then perform the physical action necessary for the manipulation. The system, in turn, should translate the physical movement and/or sensor data into an input request, update itself by changing game state and present the updated state to the user. The combination of feedback and clear interaction with the system make for a smooth and engaging experience. A user playing Jump Beat remarked on the lack of visual feedback on their jumps while another user wanted more ``visual flourishes" added to the interface to make it even more game like. One user found See-Saw boring and was uncertain about how to play if it required making their own rules. Two other users found the lack of rules interesting and wanted to continue challenging themselves but wanted their running score displayed on screen in addition to the score for each level. One See-Saw user wanted ``more feedback" and ``more points" and said the bright colors helped them ``focus on the screen."

\subsubsection{Support social interactions}

An important class of influence that had impact on motivation was the opportunity for social interaction during break time. Physically active play with a social component is engaging, motivating, and desirable. It also works better for people in shared workspaces than single player physical interaction for allowing people to take shared breaks which are less disruptive since everyone in the space is engaged in a shared activity. The physical and social interactions were noted as factors that made the games more entertaining. A number of participants commented that they liked the idea of ``social physical activity" which made ``taking a break more exciting." Six Learning To Fly participants said, ``the activity encouraged human contact'' and playing with someone made it more enjoyable. Findings from the questionnaire give the average level of excitement rated by participants as 4.23 (variance of .64), signifying a high level of excitement while doing a physical action together with others. For games with a visible display screen, spectating is an important element of the experience allowing for onlookers to transition into players and back seamlessly. We believe this ease of movement between spectatorship and playing and the overall social aspect of playing together can be a strong motivator for people to engage in physical activity more frequently.

\subsubsection{Consider the practical constraints of users' workplace culture, space, interests, and ability.
lifestyles}

81.8\% of the users felt comfortable using Glass and the heart rate sensor and some remarked that it made them ``look cool" and liked having a "private screen." Furthermore, 90.9\% of the players found the hands-free interaction enjoyable especially in Jump Beat and Learning to Fly. One user said they felt ``much more in the game" and it is ``more fun when it's on your head" about playing on Glass. Four users felt self-conscious while jumping alone in Jump Beat but all eight felt comfortable while flapping their arms together with another person in Learning to Fly, regardless of the type of work space. Though mobile phone use, especially talking and ringing in public spaces has blurred the boundaries between public and private behaviors \cite{Wei}, the same is not true of digital activity in the work place. Therefore, an understanding between people sharing office space regarding acceptable behaviors would be an important design consideration for a digital system that requires physical actions as input. Another aspect of physical interaction in a shared workspace is social embarrassment which may be overcome by creating systems that invite participation from others instead of putting one individual on display or creating systems for shared break rooms. Those participants who played Jump Beat in a shared office were conscious of their jumping even though their officemates ignored their jumping. Those who attempted Jump Beat in a private office or Learning to Fly in an open shared workspace felt comfortable doing the physical actions and did not think they were bothering anyone by their activity. Different work culture may allow for different acceptable behaviors for e.g. in an academic research lab it may be acceptable to jump or do pushups in an open work space which may not be the case in a formal corporate environment unless the individual has a private office. Similarly, people dress casually for formally depending on their work place and system design needs to take that into consideration along with target audience specifics like physical activity willingness and ability while designing interactions.

\section{Conclusion}

Based on the response to our set of physically interactive games we are excited about their potential to help combat the health hazards of sitting for long stretches. Though physical interactions and play were reported as exciting and fun, they are potentially disruptive in shared offices which could severely limit their regular use. We believe there is room for physical games in the work environment, especially social play in common areas or solo play on mobile devices like Glass for outdoors as well as private office spaces. What pleasantly surprised us was the extent to which some participants wanted to continue playing beyond the duration of a ``short" break or start playing before it was time for a break which supports our belief that intense blasts of engaging play can motivate people into being physically active. Though qualitative research can have less statistical power than quantitative research when it comes to discovering and verifying trends, our small user study has given us broad guidelines for the design of other such games. Using that information, we can design newer prototypes that can be deployed to a larger set of people for gathering more data for assessing impact on sedentary lifestyles.

\bibliographystyle{acm-sigchi}
\bibliography{activ8}

\begin{thebibliography}{10}

\bibitem{Anderson:2013:YEM:2501988.2502045}
Anderson, F., Grossman, T., Matejka, J., and Fitzmaurice, G.
\newblock Youmove: Enhancing movement training with an augmented reality
  mirror.
\newblock In {\em UIST} (2013), 311--320.

\bibitem{aspinall2013urban}
Aspinall, P., Mavros, P., Coyne, R., and Roe, J.
\newblock The urban brain: analysing outdoor physical activity with mobile eeg.
\newblock {\em British journal of sports medicine\/} (2013), bjsports--2012.

\bibitem{Bianchi}
Bianchi-Berthouze, N., Kim, W., and Patel, D.
\newblock Body movement engage you more in digital game play? and why?
\newblock In {\em Affective Comp. and Intelligent Interaction} (2007),
  102--113.

\bibitem{Chen:2012:KAW:2380116.2380171}
Chen, J., Izadi, S., and Fitzgibbon, A.
\newblock Kin\^{E}tre: Animating the world with the human body.
\newblock In {\em UIST}, ACM (2012), 435--444.

\bibitem{Colaco:2013:MCL:2501988.2502042}
Cola\c{c}o, A., Kirmani, A., Yang, H.~S., Gong, N.-W., Schmandt, C., and Goyal,
  V.~K.
\newblock Mime: Compact, low power 3d gesture sensing for interaction with head
  mounted displays.
\newblock In {\em UIST} (2013), 227--236.

\bibitem{Dantzig:2013:TPM:2559051.2559079}
Dantzig, S., Geleijnse, G., and Halteren, A.~T.
\newblock Toward a persuasive mobile application to reduce sedentary behavior.
\newblock {\em Personal Ubiquitous Comput. 17}, 6 (Aug. 2013), 1237--1246.

\bibitem{deOliveira:2008:TEE:1409240.1409268}
de~Oliveira, R., and Oliver, N.
\newblock Triplebeat: Enhancing exercise performance with persuasion.
\newblock In {\em Proc. MobileHCI}, ACM (2008), 255--264.

\bibitem{deterding2011game}
Deterding, S., Dixon, D., Khaled, R., and Nacke, L.
\newblock From game design elements to gamefulness: defining gamification.
\newblock In {\em Proceedings of the 15th International Academic MindTrek
  Conference: Envisioning Future Media Environments}, ACM (2011), 9--15.

\bibitem{Fujiki:2010:IPA:1753846.1754146}
Fujiki, Y.
\newblock iphone as a physical activity measurement platform.
\newblock In {\em CHI EA} (2010), 4315--4320.

\bibitem{Gao}
Gao, Z., Zhang, T., and Stodden, D.
\newblock Children's physical activity levels and psychological correlates in
  interactive dance versus aerobic dance.
\newblock In {\em J Sport Health Sci.} (2013), 146--151.

\bibitem{Hamilton}
Hamilton, M.~T., Healy, G.~N., Dunstan, D.~W., Zderic, T.~W., and N., O.
\newblock Too little exercise and too much sitting: Inactivity physiology and
  the need for new recommendations on sedentary behavior.
\newblock In {\em Curr Cardiovasc Risk Rep.} (2008), 292--298.

\bibitem{Healy}
Healy, G.~N., Dunstan, D.~W., Salmon, J., Cerin, E., Shaw, J.~E., Zimmet,
  P.~Z., and Owen, N.
\newblock Breaks in sedentary time beneficial associations with metabolic risk.
\newblock {\em Diabetes care 31}, 4 (2008), 661--666.

\bibitem{Hennings}
Henning, R., Jacques, P., Kissel, G., Sullivan, A., and Alteras-Webb, S.
\newblock Frequent short rest breaks from computer work: effects on
  productivity and well-being at two field sites.
\newblock In {\em Ergonomics} (1997), 78--91.

\bibitem{Consolvo}
Klasjna, P., Consolvo, S., McDonald, D., Landay, J., and Pratt, W.
\newblock Using mobile \& personal sensing technologies to support health
  behavior change in everyday life: lessons learned.
\newblock In {\em Amer. Med. Info. Assoc.} (2009).

\bibitem{Lee}
Lee, I.~M., Shiroma, E.~J., Lobelo, F., Puska, P., Blair, S.~N., and
  Katzmarzyk, P.~T.
\newblock Effect of physical inactivity on major non-communicable diseases
  worldwide: an analysis of burden of disease and life expectancy.
\newblock {\em The Lancet 380}, 9838 (2012), 219--229.

\bibitem{Lin:2006:FEP:2166283.2166299}
Lin, J.~J., Mamykina, L., Lindtner, S., Delajoux, G., and Strub, H.~B.
\newblock Fish'n'steps: Encouraging physical activity with an interactive
  computer game.
\newblock In {\em Proc. UBICOMP} (2006), 261--278.

\bibitem{Lyons}
Lyons, E., D.F., T., Ward, D., Bowling, J., Ribisl, K., and Kalyararaman, S.
\newblock Energy expenditure and enjoyment during video game play: differences
  by game type.
\newblock In {\em Med Sci Sports Exerc.}, vol.~43 (2011), 1987--93.

\bibitem{Morris:2008:SUI:1357054.1357337}
Morris, D., Brush, A.~B., and Meyers, B.~R.
\newblock Superbreak: Using interactivity to enhance ergonomic typing breaks.
\newblock In {\em Proc. CHI} (2008), 1817--1826.

\bibitem{Mueller}
Mueller, F., and Agamanolis, S.
\newblock Sports over a distance.
\newblock In {\em CIE} (2005), 3--4.

\bibitem{Edge}
Mueller, F., Edge, D., Vetere, F., Gibbs, M., Agamanolis, S., Bongers, B., and
  Sheridan, J.
\newblock Designing sports: A framework for exertion games.
\newblock In {\em Proc. CHI} (2011), 2651--2660.

\bibitem{Isbister}
Mueller, F., and Isbister, K.
\newblock Movement-based game guidelines.
\newblock In {\em Proceedings of the 32nd annual ACM conference on Human
  factors in computing systems}, ACM (2014), 2191--2200.

\bibitem{Ross}
Ross, R., and McGuire, A.
\newblock Incidental physical activity is positively associated with
  cardiorespiratory fitness.
\newblock {\em Medicine \& Science in Sports \& Exercise 44}, 3 (2012),
  2189--2194.

\bibitem{Wei}
Wei, R., and Leung, L.
\newblock Blurring public and private behaviors in public space: policy
  challenges in the use and improper use of the cell phone.
\newblock {\em Telematics and Informatics 16}, 1-2 (1999), 11--26.

\bibitem{mit.00210851220110101}
Wigdor, D., and Wixon, D.
\newblock {\em Brave NUI world [electronic resource] : designing natural user
  interfaces for touch and gesture / Daniel Wigdor, Dennis Wixon.}
\newblock Burlington, Mass. : Morgan Kaufmann, c2011 (Norwood, Mass. :
  Books24x7.com), 2011.

\bibitem{Yim:2007:UGI:1328202.1328232}
Yim, J., and Graham, T. C.~N.
\newblock Using games to increase exercise motivation.
\newblock In {\em Proc. Future Play}, ACM (2007), 166--173.

\end{thebibliography}

\end{document}